# TUNE EVALUATION FROM PHASED BPM TURN-BY-TURN DATA*

Y. Alexahin, E. Gianfelice-Wendt, W. Marsh, FNAL, Batavia, IL 60510, U.S.A.

*Abstract*

In fast ramping synchrotrons like the Fermilab Booster the conventional methods of betatron tune evaluation from the turn-by-turn data may not work due to rapid changes of the tunes (sometimes in a course of a few dozens of turns) and a high level of noise. We propose a technique based on phasing of signals from a large number of BPMs which significantly increases the signal to noise ratio. Implementation of the method in the Fermilab Booster control system is described and some measurement results are presented.

## INTRODUCTION

Turn-by-turn beam position measurements still remain the most reliable tool for tune determination. Standard method for tune evaluation – FFT – has resolution $\sim 1/N$, $N$ being the number of turns, which is insufficient in the case of rapid changes of the tunes and/or fast decoherence of the betatron oscillations.

Much better precision can be achieved with the so-called Continuous Fourier Transform (CFT) method [1] which consists in evaluation of the sum

$$X(\nu) = \frac{1}{N}\sum_{n=1}^{N} e^{-2\pi i \nu (n-1)} x_n \quad (1)$$

as a function of continuous variable $\nu$ and finding the maximum of $|X(\nu)|$.

In absence of random noise CFT provides precision $\sim 1/N^2$. There are methods – i.e. the Hanning windowing technique [1] – which further improve precision, up to $1/N^4$, but they fail in the presence of noise.

In the case of white noise the CFT provides tune with the r.m.s. error [2]

$$\sigma_\nu \approx \frac{\sqrt{6}\sigma}{\pi N^{3/2} a}, \quad (2)$$

where $\sigma$ is the r.m.s. value of BPM errors and $a$ is the betatron oscillations amplitude, which is better than the FFT error even in absence of noise. But it may be still not enough in a situation when the noise level is high and only a small number of turns is available.

In this report we show how the *a priori* knowledge of machine optics may help to drastically improve the precision of tune determination.

## BASIC IDEA

Let us start with a real-life example. Figure 1 top shows a 32-turns CFT spectrum obtained from one of the best horizontal BPMs (B:HST14S) in the Fermilab Booster at ~24.5 ms into the ramp. The beam energy at this time is

___________________________________________
* Work supported by Fermi Research Alliance, LLC under Contract DE-AC02-07CH11359 with the U.S. DOE.
#alexahin@fnal.gov

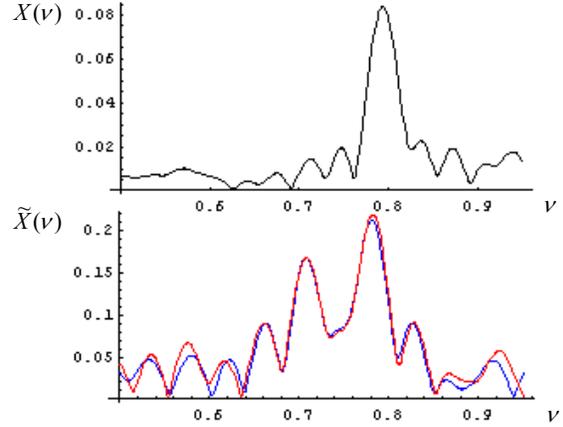

Figure 1 (color): CFT spectrum from a single horizontal BPM (top) and combined CFT spectrum from 24 BPMs (bottom) obtained with two versions of the method.

already quite high and the horizontal kicker power is not sufficient to excite noticeable oscillations. Only one mode can be seen (presumably vertical) due to self-excitation. It is not possible to extract information from a single BPM data on the other mode since it is completely suppressed by the strong self-excited mode and the noise.

We may try, however, to use information from all available BPMs in assumption that the betatron phase advance between them does not differ too much from theoretical values $\varphi^{(k)}_{x,y}$, $k$ being the BPM index. In this case the oscillations propagate around the ring as

$$x_n^{(k)} = \frac{1}{2} a_x^{(k)} \exp[2\pi i Q_x (n-1) + i\varphi_x^{(k)} + i\psi_{x0}] + c.c. \quad (3)$$

where $Q_x$ is the betatron tune and $\psi_{x0}$ is the initial phase (we will write all formulas for the horizontal plane only). Amplitude $a_x^{(k)}$ varies from BPM to BPM according to the betatron function value:

$$a_x^{(k)} = \sqrt{\beta_x^{(k)} E_x}, \quad (4)$$

$E_x$ being the Courant-Snyder invariant of oscillations.

When looking for the horizontal tune we may use data from all horizontal BPMs to construct a phased sum

$$\tilde{x}_n = \sum_k w_k x_n^{(k)} \exp[-i\varphi_x^{(k)}] \quad (5)$$

for each turn $n$, where $w_k$ are some weights, and perform the CFT analysis using $\tilde{x}_n$. Weights $w_k$ may reflect the quality of individual BPM data, here we set $w_k = 1$.

From eqs. (1), (3) and (5) one can easily see that the proper part of the signal propagating with expected phase advance is amplified by a factor of $N_{BPM}$ whereas the alien modes and random noise are amplified only as $\sqrt{N_{BPM}}$ so that the signal to noise ratio is improved by a factor of $\sqrt{N_{BPM}}$.

Figure 1 bottom shows with blue line the spectrum of horizontal oscillations obtained with this method using all 24 Booster horizontal BPMs at high $\beta_x$ locations.

One can see a peak which appeared close to the theoretical value of the horizontal tune $Q_x^{(theo)} = 6.7$ which was almost completely suppressed in a single BPM spectrum (Fig. 1 top). However, such closeness to the theoretical value may raise a suspicion that the observed peak is an artifact of the method since this value is embedded in the theoretical phase advance distribution.

*Variable Reference Tune*

We can modify the algorithm so that there was no preferred value of the tune. Specifically, when performing the CFT we may assume that theoretical tune is equal to $\nu$ and spread the phase advance difference uniformly around the ring

$$\varphi_x^{(k)} = \varphi_{x0}^{(k)} + s_k(\nu - Q_{x0}) \qquad (6)$$

where $s_k$ is the $k$-th BPM longitudinal position. The final formula for the phased CFT will now look as

$$\widetilde{X}(\nu) = \frac{1}{N}\sum_{n=1}^{N} e^{-2\pi i \nu(n-1)} \sum_k w_k x_n^{(k)} e^{-i\varphi_x^{(k)} - is_k(\nu - Q_{x0})} \qquad (7)$$

Spectrum obtained with this formula is shown in Fig. 1 bottom with red line which confirms the validity of the previous result.

When using eq. (7) the total of $N_{BPM} \times N$ terms have to be calculated for each $\nu$ value, while in the original formulation the phased sums (5) were calculated just once and then the CFT for each $\nu$ required calculation of only $N$ terms. Our example shows that the complication which variation of the reference tune presents is not justified, at least for the Fermilab Booster.

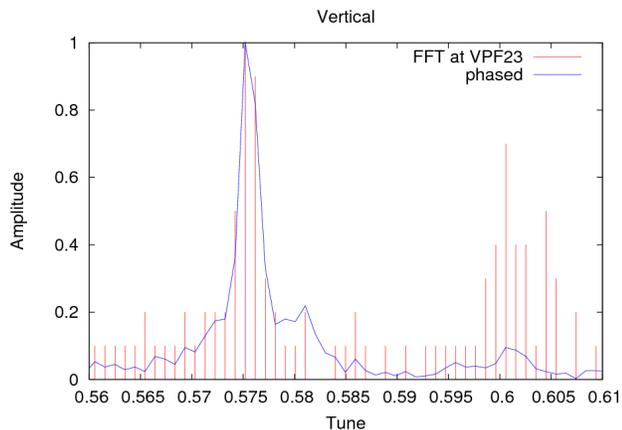

Figure 2 (color): FFT of TBT data from one BPM (red) and of the phased sum of all 118 Tevatron vertical BPMs (blue).

## TEVATRON TUNES AT ENERGY RAMP

Control of betatron tunes during acceleration is essential for good performance of the Tevatron but presents significant difficulties due to relatively fast variation of parameters, high chromaticity and strong noise. Also, due to long turn-around time it is desirable to determine both tunes in a single measurement pinging the beam with one (e.g. horizontal) kicker and relying on residual coupling for excitation of oscillations in the other (vertical) plane.

However, using information from individual BPMs this was not always possible. Figure 2 shows with red bars a 1024 turns vertical FFT spectrum obtained from a single BPM after a horizontal ping at 400 GeV. The only prominent peak (besides horizontal tune at $\nu = 0.575$) is a fake line at $\nu = 0.6$ produced by BPM electronics. The vertical tune line is drowned in noise. Spectra from other BPMs look similar.

The blue line in Fig. 2 shows the FFT of a phased sum for 63 vertical BPMs that worked at the time of the measurements. One can see the noise to be strongly suppressed and the vertical line at $\nu = 0.58$ to become second highest. However, the relative height of the peaks remained approximately the same since the difference in phase advances between the two modes is very small.

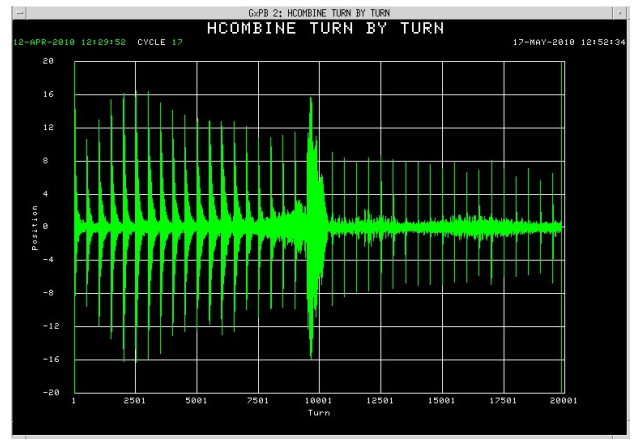

Figure 3: Combined Horizontal BPM data with running average subtracted

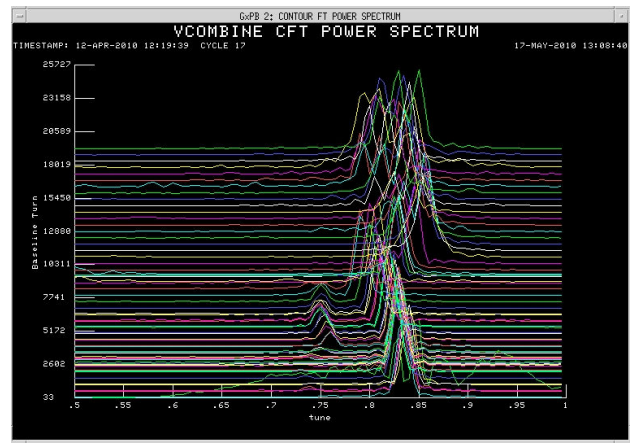

Figure 4 (color): Mountain range plot of the vertically pinged CFT spectra throughout the Booster cycle.

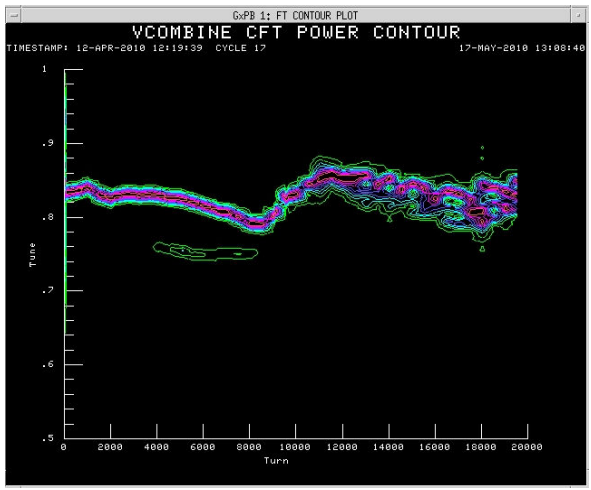

Figure 5 (color): Contour plot of the combined CFT spectra of the vertically pinged beam showing the tunes evolution through the Booster cycle.

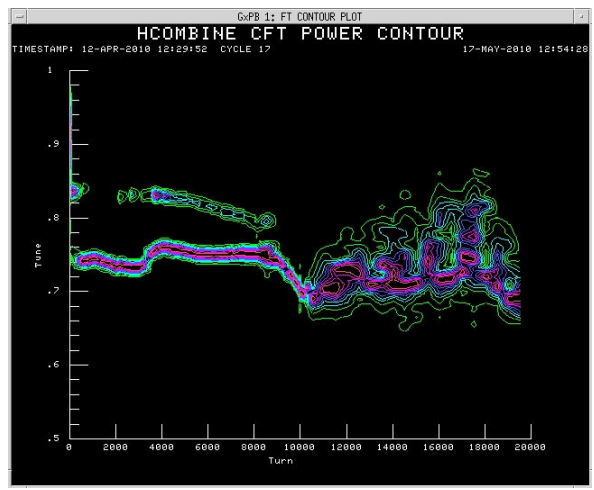

Figure 6 (color): Contour plot of the combined CFT spectra of the horizontally pinged beam.

## BOOSTER CONSOLE APPLICATION

The algorithm was implemented in an ACNET control system application to give operators on-line feedback on the Booster tunes. The application – B38 – existed before but failed to provide information on the horizontal tune for the most part of the ramp using information from individual BPMs.

The application works as follows. A kicker – horizontal or vertical – is set up to kick the beam every 500 turns. On completion of the ramp the application reads out the turn by turn BPM data for all turns and all BPM's. The horizontal or vertical BPM's readings are then combined for each turn according to eq. (5) and the moments of pings on the beam are identified by the oscillation onset from a running average (Fig. 3). Continuous Fourier Transforms are then performed for each ping. Each spectrum is normalized so that the peak value is the same for all pings.

The spectra can be viewed as a mountain range plot or a contour plot shown in Figs. 4 and 5 respectively for the case of vertical oscillations.

As already mentioned, determination of the horizontal tune presents most difficulties in the Booster. Figure 6 shows a contour plot of the CFT spectrum obtained from a phased sum (5) while Fig. 7 – for comparison – shows spectrum from only one BPM.

The achieved clarification of the spectra allowed better tuning of the Booster which resulted in noticeably improved performance. The remaining fuzziness in the second part of the ramp is a result of systematic noise probably produced by BPM electronics.

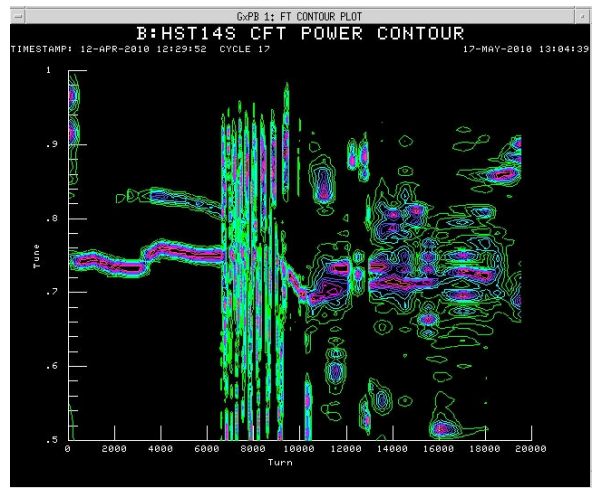

Figure 7 (color): Contour plot of the CFT spectra of the horizontally pinged beam obtained from one BPM.

## ACKNOWLEDGEMENTS

The authors are grateful to B. Pellico and K. Triplett for many rounds of the Booster TBT data taking which were very helpful for the development of this method.

## REFERENCES

[1] R. Bartolini et al., Particle Accelerators, **56**, 167-199 (1996).

[2] Y. Alexahin, E. Gianfelice-Wendt, FERMILAB-PUB-06-093-AD (2006).